\def\ave#1{\left\langle #1 \right\rangle}

\def\arcminf {\hbox{$.\!\!^{\prime}$}}

\documentclass[useAMS,usenatbib]{mn2e}

\usepackage{graphicx}
\usepackage{epsfig}  
\usepackage{times}    
\usepackage{subfigure}
\title[Cosmology with the cluster mass function]{Cosmology with the cluster mass function: mass estimators and shape systematics in large weak lensing surveys}
\author[V. L. Corless \& L. J. King]{Virginia L. Corless$^{1}$\thanks{E-mail:
vc258@ast.cam.ac.uk} and Lindsay J. King$^{1}$\\
$^{1}$Institute of Astronomy, University of Cambridge, Madingley Road, Cambridge, United Kingdom}

\begin{document}

\date{2008 August 18}

\pagerange{\pageref{firstpage}--\pageref{lastpage}} \pubyear{2008}

\maketitle

\label{firstpage}

\begin{abstract}
Accurate measurement of the cluster mass function is a crucial element in efforts to constrain structure formation models, the normalisation of the matter power spectrum and the cosmological matter density, and the nature and evolution of dark energy.  Large weak lensing surveys of $\sim 20,000$ galaxy clusters and groups will be key tools in the observational pursuit of that goal; first-generation surveys such as those using the Sloan Digital Sky Survey are already beginning to calibrate previous measurements of the mass function from X-ray observations and to extend existing constraints down to galaxy group scales.  These weak lensing studies proceed by stacking the lensing signals of many clusters and groups binned by mass-correlated observables such as richness and luminosity; typically such analyses ignore the triaxial structure of dark matter halos on the assumption that the averaging of many shear signals within each mass bin makes its effects (as large as factors of two in mass model parameter estimates from individual clusters) negligible.  We test this assumption utilizing a population of 15,000 analytic triaxial dark matter halos spanning two dex in mass, and find that triaxiality can bias 3D virial mass estimates compared to those for a spherical population by a few percent if suboptimal mass estimators are used.  This bias affects not only direct lensing constraints on the mass function but can also affect the scatter and normalization of the mass-observable relations derived from lensing that are so crucial to constraining the cluster mass function with large samples.  However, we demonstrate that a careful choice of mass estimator can remove the bias very effectively if the lensing signals from a sufficient number of triaxial halos are averaged together, and further quantify that sufficient number for adequate shape averaging.  We thus show that by choosing observable bins to contain an adequate number of halos and by utilizing a carefully chosen 3D mass estimator stacked weak-lensing analyses can give unbiased constraints on the triaxial mass function.
\end{abstract}



\section{Introduction}
The number of galaxy clusters and groups present in the universe as a function of mass and redshift is a strong constraint on cosmological models:  measuring this cluster mass function over a wide range of masses and redshifts is a crucial element in the effort to constrain structure formation models, cosmological parameters $\Omega_M$ and $\sigma_8$, and the nature and evolution of dark energy.  The mass function has been well studied in simulations (e.g. \cite{evrard02}) and in nature in the X-ray (e.g. \cite{vikhlinin08}; \cite{reiprich02}), the optical (e.g. \cite{bahcall03}), and in first-generation weak lensing studies in large surveys such as the Sloan Digital Sky Survey (SDSS, \cite{york00}; \cite{johnston08}).  A detailed review of the theory of the cluster mass function and its observational pursuit is given by \cite{voit04}.

\begin{figure*}
\centering
\rotatebox{0}{
\resizebox{14cm}{!}
{\includegraphics{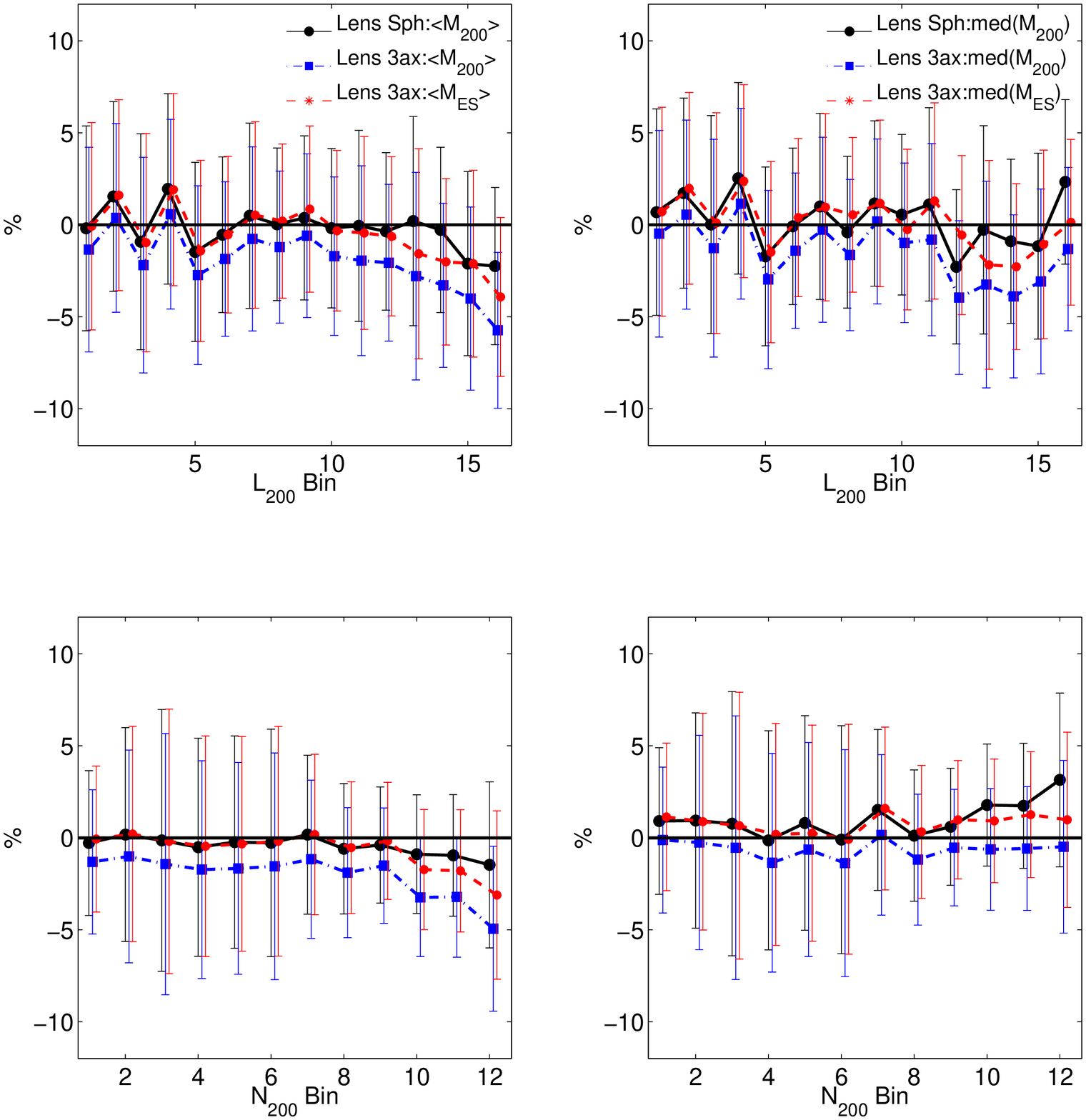}}
}
 \caption{Offsets in masses recovered, averaged across 50 noise realizations, for lensing by identical populations of 15,000 spherical and triaxial lensing halos, binned using SDSS observables.  The top left panel shows the percent offset of the mean recovered lensing mass compared to the {\it true} mean mass $\ave M_{200}$ for each luminosity $L_{200}$ bin.  The top right panel is the same but compares with the {\it true} median mass med$(M_{200})$.  The bottom panels show the same for the richness $N_{200}$ bins.  The solid black lines plot the results for lensing by a population of spherical halos, the blue dot-dash lines show them for lensing by the corresponding triaxial halo population, and the red dashed line shows the results for lensing by the triaxial population compared to the true mean and median effective spherical mass $M_{ES}$ in each bin.  The effective spherical mass $M_{ES}$ is the triaxial mass measure most closely recovered by a stacked lensing analysis.  The error bars give the standard deviation across the 50 realizations.}
 \label{fig:plot1}
\end{figure*}

Weak lensing is of growing importance in the effort to constrain the mass function as the next generation of weak lensing surveys promise much larger samples of galaxy clusters and groups over a wider range of redshifts than have been studied to date.  It may be a preferred method because the mass measurements it provides do not depend on baryonic physics and are largely independent of the observables by which lens systems are sorted into mass bins (X-ray temperature or luminosity; optical luminosity or richness), making it less likely to be biased by systematic errors specific to one type of observational data (see e.g. \cite{mandelbaum07}).  However, lensing is subject to its own systematic and statistical errors, which must be controlled for it to become reliable as a measure of the mass function.  The Dark Energy Task Force (DETF) sets the goal that the RMS error in the mean and variance of the mass estimates in current and next-generation weak lensing surveys be reduced to no more than $14\%$ and $11\%$ respectively, and ideally to less than $2\%$ (\cite{albrecht06}) in order to accurately construct a mass function.  A full understanding of all sources of error is crucial to achieve those levels.  Some possible sources are unaccounted-for cluster substructure, scatter from large scale line-of-sight structures (\cite{dodelson04}; \cite{hoekstra03}), incomplete cluster and group samples, and mischaracterization of the scatter and normalization in the mass-observable relations used to bin and stack the weak lensing signals.  

In this paper we address one potentially important systematic error: the neglect of the triaxial structure of the dark matter halos that shape clusters and groups.  CDM structure formation simulations such as those of \cite{shaw06} and \cite{bett07} indicate populations of significantly triaxial halos that are more prolate than oblate with minor axis ratios as small as 0.3.  Several studies have examined the importance of this triaxial structure in lensing analyses of individual clusters, which typically employ spherical, or at best elliptical, models (\cite{oguri05}; \cite{corless07}; \cite{corless08}; \cite{meneghetti07}), and found its neglect to cause errors of up to factors of two in estimates of virial mass and halo concentration.
\begin{figure}
\centering
\rotatebox{0}{
\resizebox{7cm}{!}
{\includegraphics{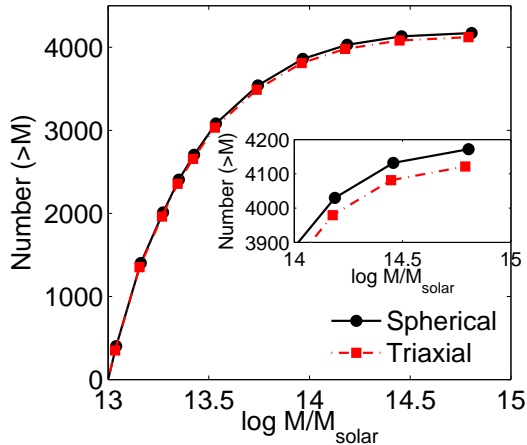}}
}
 \caption{The mean cumulative mass function for halos of $M > 1.0\times 10^{13}M_{\odot}$ recovered from the 50 noise realizations of Figure \ref{fig:plot1}, for lensing by identical populations of 15,000 spherical and triaxial lensing halos, binned using the SDSS richness observable $N_{200}$.  At low masses the impact of the triaxiality bias is very small; at high masses, the region richest in information about cosmological parameters, it is small but potentially significant.}
 \label{fig:plot1p5}
\end{figure}

Typically triaxiality is neglected in stacked weak lensing analyses measuring halo masses because it is argued that the averaging of the lensing shear signals of many halos randomly oriented in each bin makes the spherical mass derived from the mean signal an accurate estimator of the true bin mean mass (see e.g. \cite{johnston08}).  However, Corless $\&$ King (2007; 2008) showed that while indeed the mean mass across a population of triaxial halos fit individually with spherical models is a good estimator of the mean population value, the scatter is asymmetric and the error estimates incorrect.  Given this, the interpretation of the mean shear signal in a bin stretched across a steeply sloped mass function is complex and not obvious.  Neglecting this complexity in mass modelling may lead to mass estimates that cannot be interpreted as theoretical quantities such as the virial mass, an interpretation which is necessary to make comparisons between observational results and theoretical predictions.  The importance of the mass function for cosmology and the increasing efforts that will be devoted to constraining it in future lens surveys make it crucially important to fully understand any impact triaxiality may have on the accurate and clear measurement of galaxy cluster and group masses.

To this end, in this paper we directly model the impacts of triaxiality on the masses recovered from weak lensing surveys, using an analytic lens population based on an SDSS sample. Section \ref{sec:sample} describes our simulations of lensing by a large population of triaxial galaxy clusters and groups, Section \ref{sec:lens} outlines the mass-observable binning and simple stacked weak lensing techniques we employ to fit their masses, and Section \ref{sec:results} presents our results and some discussion.

\begin{figure*}
\centering
\rotatebox{0}{
\resizebox{14cm}{!}
{\includegraphics{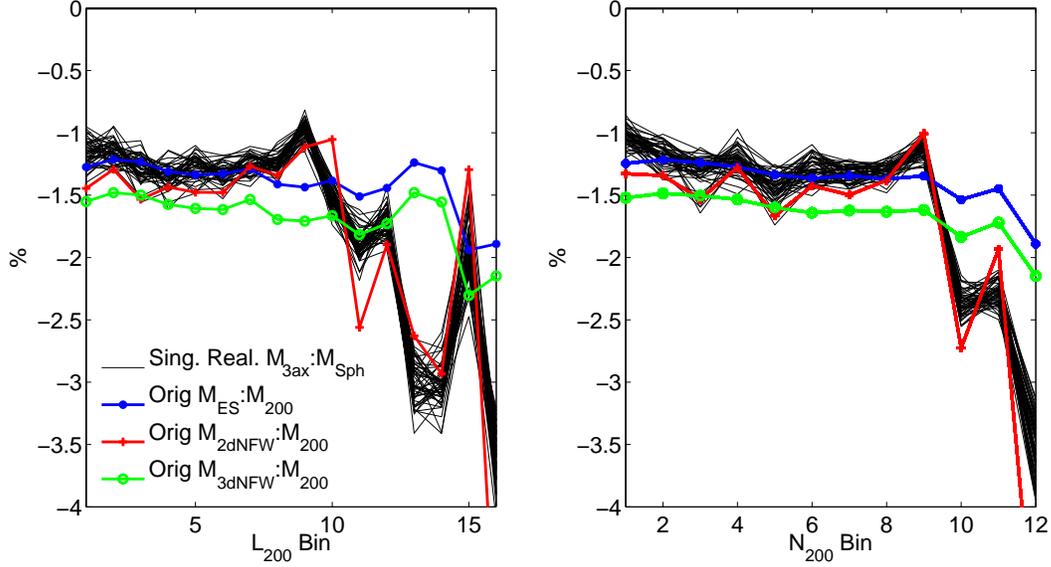}}
}
 \caption{Offsets in the masses recovered, averaged across 50 noise realizations, for lensing by identical populations of 15,000 spherical and triaxial lensing halos, binned using SDSS observables.  The black lines plot the percent offset of the lensing mass recovered from the triaxial lensing population compared to that for the spherical population in each luminosity $L_{200}$ (left-hand panel) and richness $N_{200}$ (right-hand panel) bin, for all 50 realizations.  Overplotted are comparisons of the true mean values of the various mass estimators discussed in Sec. \ref{sec:mass} to the true mean $M_{200}$ value for the triaxial population in each bin (all mass estimators are equal for the spherical population).  $M_{ES}$ is plotted with blue stars, $M_{2DNFW}$ with red crosses, and $M_{3DNFW}$ with green circles.  The effective spherical mass $M_{ES}$ is the triaxial mass measure most closely recovered by a stacked lensing analysis in low mass bins that contain many halos; the 2D lensing mass $M_{2DNFW}$ performs best in the high mass bins with few halos where shape averaging is incomplete.}
 \label{fig:plot2}
\end{figure*}

\section{Simulations: Lensing through a cosmological population of triaxial halos}\label{sec:sample}
\subsection{Simulated cluster sample}\label{sec:sims}
To begin, we generate a catalogue of triaxial dark matter halos with a generalised NFW density profile
\begin{equation}
\rho(R) = \frac{\delta_c \rho_c(z)}{R/R_s(1 + R/R_s)^2}
\label{eq:3axrho}
\end{equation}
where $\delta_c$ is the characteristic overdensity of the halo, $\rho_c$ the critical density of the universe at the redshift $z$ of the cluster, $R_s$ a scale radius, $R$ a triaxial radius 
\begin{equation}
R^2 = \frac{X^2}{a^2} + \frac{Y^2}{b^2} + \frac{Z^2}{c^2},\textrm{         }(a\leq  b \leq  c = 1),\label{eq:3axR}\end{equation}
and $a/c$ and $b/c$ the minor:major and intermediate:major axis ratios, respectively.  The NFW concentration is defined as usual
\begin{equation}C = \frac{R_{200}}{R_s},\label{eq:3axC}\end{equation}
and we employ a fully triaxial virial mass
\begin{equation}M_{200} = \frac{800\pi}{3}abR_{200}^3\rho_c,\label{eq:3axM200}\end{equation} where $R_{200}$ is the triaxial radius of the triaxial isodensity surface within which the mean density is 200 times the critical density.  We choose to define the virial mass as a fully triaxial quantity because such a definition is well-supported by collapse theory (\cite{sheth01}) and is a more realistic representation of dark matter halo shapes.

We assemble a physically representative population of 15,000 halos as follows
\begin{itemize}
\item Draw a mass $M_{200}$ for each halo from the mass function of \cite{evrard02} at $z=0.25$, using the best fit parameters under the $\Lambda$CDM model for the differential number density of dark matter halos as a function of mass and redshift
\begin{equation}n(M,z) = \frac{A\bar{\rho}_m(z)}{M}\alpha_{eff}(M)\exp\left[-|\ln\sigma^{-1}(M)+B|^{\epsilon}\right]\end{equation}
where $\alpha_{eff}$ is the effective logarithmic slope and $\sigma^2(M)$ is the variance of the density field smoothed on scales enclosing mass $M$ at the mean density $\bar{\rho}_m(z)$.  We choose halos that range in mass from $10^{12.5}$M$_{\odot}$ to $10^{15.5}$M$_{\odot}$ in order to obtain a complete sample over the range of $10^{13} - 10^{15}$ M$_{\odot}$ when potential scatter from lensing shape dispersion and triaxiality are accounted for.
\item Assign a concentration $C$ to each halo according to the mass-concentration relationship for relaxed halos at $z=0.25$ of \cite{neto07}, derived from the simulated groups and clusters in the Millennium simulation  
\begin{equation}C = \frac{5.26}{1+z}\left(\frac{M_{200}}{10^{14} h^{-1} M_{\odot}}\right)^{-0.1},\label{eq:M-C}\end{equation}
with a log-normal scatter
\begin{equation}p(\log C) = \frac{1}{\sigma \sqrt{2\pi}}\exp\left[\frac{1}{2}\left(\frac{\log C - {\ave {\log C}}}{\sigma}\right)^2\right]\end{equation}
where $\ave{\log C}$ is calculated via Eq. \ref{eq:M-C} for a given $M_{200}$ and the dispersion $\sigma$ is taken to be 0.09 for masses less than $10^{15}$ M$_{\odot}$ and 0.06 for masses greater than that threshold, in rough agreement with the results of \cite{neto07}.
\item Assign each halo a triaxial shape defined by  minor and intermediate axis ratios $a$ and $b$ and two randomly distributed orientation angles.  The axis ratios are drawn from the distributions found in the structure formation simulations of \cite{shaw06}.
\item Assign each halo a luminosity and richness drawn from the mass-richness ($N_{200}$: galaxy number) and mass-luminosity ($L_{200}$: $i$-band luminosity) relations from the stacked weak lensing analysis of the SDSS of \cite{johnston08}:
\begin{equation}N_{200}(M_{200}) = 20\left(\frac{M_{200}}{M_{200|20}}\right)^{1/\alpha_N}\label{eq:M-N}\end{equation}
with
\begin{eqnarray}M_{200|20} = (8.8\pm 0.4_{stat}\pm1.1_{sys})&\times& 10^{13}h^{-1}M_{\odot}\nonumber\\
\alpha_N&=&1.28\pm 0.04\nonumber\end{eqnarray}
and
\begin{equation}L_{200}(M_{200})= 40\left(\frac{M_{200}}{M_{200|40}}\right)^{1/\alpha_L}10^{10}h^{-2}L_{\odot}\label{eq:M-L}\end{equation}
with
\begin{eqnarray}M_{200|40} = (9.5\pm 0.4_{stat}\pm1.2_{sys})&\times& 10^{13}h^{-1}M_{\odot}.\nonumber\\
\alpha_L&=&1.22\pm 0.04\nonumber\end{eqnarray}
There is an intrinsic scatter in these mass-observable relations which is not well constrained but expected to be of order $\sim 20\%$.  In this study, because we seek to isolate the effects of triaxiality, we neglect this scatter and use the best fit values of the mass-richness and mass-luminosity parameters exactly with no error or scatter.
\end{itemize}

For every case in this paper, an additional identical population of spherical halos with the same $M_{200}$, $C$, $N_{200}$, and $L_{200}$ values is assembled for comparison.

\subsection{Measures of Mass}\label{sec:mass}
Above, we defined $M_{200}$ as a triaxial quantity, a choice we prefer because it describes halos more as we expect them to really be.  However, many other mass measures have been used in previous work, and the choice of measure can be important, as we will demonstrate later in this paper.  Three alternatives for describing a triaxial lensing halo's mass are
\begin{itemize}
\item An effective spherical mass and concentration where the effective spherical virial radius $r_{200, ES}$ is defined as the radius of the spherical surface (which is {\it not} an isodensity surface) within which the mean density is 200 times the critical density.  The effective spherical virial mass is the mass within that sphere
\begin{equation}M_{ES} = (800\pi /3) \rho_c r_{200, ES}^3,\label{eq:MES}\end{equation}
and the effective spherical concentration $C_{ES}$ is the ratio of $r_{200, ES}$ to the geometric mean of the triaxial scale radii $C_{ES} = r_{200, ES}/r_{s, ES}$, where $r_{s, ES} = R_s(abc)^{1/3}$.  $M_{ES}$ is always less than $M_{200}$, usually by only a few percent.  This parameterisation is equivalent to directly calculating $r_{200,ES}$ from the raw mass distribution of a cluster in a structure formation simulation.  This mass estimator is very similar to that adopted by \cite{jing02} to describe triaxial dark matter halo masses.  $M_{ES}$ is better defined than $C_{ES}$, because the concentration is meaningful only as a parallel to the NFW profile, where as the effective spherical mass is well defined for any model.
\item An effective lensing 2D circular NFW mass, calculated by fitting the projected density profile of a spherical NFW model to the projected density profile of the triaxial halo, giving mass and concentration $M_{2DNFW}$ and $C_{2DNFW}$.  This mass makes sense only in a lensing context, as $M_{2DNFW}$ will differ depending on the direction of projection for a given triaxial halo.
\item An effective 3D spherical NFW mass, calculated by fitting the 3D density profile of a spherical NFW model to the density profile of the triaxial halo, giving mass and concentration  $M_{3DNFW}$ and $C_{3DNFW}$.   This parameterisation is equivalent to fitting a spherical NFW model to the mass distribution of a cluster in a structure formation simulation.
\end{itemize}
All of these masses are equal for spherical NFW halos.  

\subsection{Lensing Simulations}
Weak lensing distorts the shapes and number densities of background galaxies.  The shape and orientation of a background galaxy can be described by a complex ellipticity $\epsilon^s$, with modulus $|\epsilon^s|=(1-b/a)/(1+b/a)$, where $b/a$ is the minor:major axis ratio, and a phase that is twice the position angle $\phi$, $\epsilon^s=|\epsilon^s|e^{2i\phi}$.  The galaxy's shape is distorted by the complex weak lensing reduced shear, $g=\gamma/(1-\kappa)$, where $\gamma$ is the complex lensing shear and $\kappa$ the convergence, such that the ellipticity of the lensed galaxy $\epsilon$ becomes
\begin{equation}\epsilon = \frac{\epsilon^s + g}{1 + g^{\ast}\epsilon^s} \approx \epsilon^s + \gamma\label{eq:lens}\end{equation}
in the limit of weak deflections (the $*$ denotes complex conjugation).  The distributions of ellipticities for the lensed and unlensed populations are related by 
\begin{equation}p_{\epsilon} = p_{\epsilon^s}\left|\frac{d^2\epsilon^s}{d^2\epsilon}\right|;\end{equation}
assuming a zero-mean unlensed population, the expectation value for the lensed ellipticity on a piece of sky is $\ave\epsilon = g \approx \gamma$.  Thus the measured shapes of images can be used to estimate the shear profile generated by an astronomical lens. Lensing also changes the number counts of galaxies on the sky via competing effects; some faint sources in highly magnified regions are made brighter and pushed above the flux limit of the observation, but those same regions are stretched by the lensing across a larger patch of sky and so the number density of sources is reduced (\cite{canizares82}).

The lensing convergence and shear of triaxial NFW halos are calculated numerically by integrating over combinations of the convergence of a spherical NFW and its derivatives, scaled by several factors that account for the distribution of mass along the axis oriented along the line-of-sight. A full derivation of the lensing properties of triaxial NFW halos is given by \cite{oguri03}, and extended in \cite{corless07}.  

We simulate weak lensing through each of the 15,000 halos in each population, applying Eq. \ref{eq:lens} and the numerically calculated convergence and shear to lens catalogues of randomly positioned and oriented background galaxies with intrinsic shapes $\epsilon^s$ drawn from a Gaussian distribution with dispersion $\sigma=0.2$ in the modulus $|\epsilon^s|$, number density of 5/arcmin$^2$, and a redshift distribution modelled roughly on that of the galaxy population of SDSS Data Release 6 that peaks at $z=0.45$ and has a tail out to $z=0.8$ (\cite{oyaizu07}; note these are significantly lower redshifts than in a typical deep weak lensing study).  For each triaxial population, the catalogue of background galaxies is randomly generated independently for every individual lensing halo.  Background catalogues exactly matching those of their corresponding triaxial halos are used for the halos in spherical population.  Thus, we test the changes in the lensing signal and reconstruction when halos identical in all but shape lens through an identical universe.

All the lens halos are placed at the median redshift $z=0.25$ of the SDSS galaxy cluster population described in \cite{koester07}, such that for the cluster population angular and physical scales become equivalent measures and the impact of the cluster redshift distribution need to not be accounted for.  Again, this 
simplifying assumption is justified by our aim to isolate and understand the impacts of triaxiality.  Magnification effects and Poisson noise in the galaxy distribution are accounted for.  Throughout we assume a concordance cosmology with matter density parameter $\Omega_{\rm m}=0.3$, Hubble parameter $h=0.72$ in units of 100 km/s/Mpc, and a cosmological constant $\Omega_{\Lambda} = 0.7$.

\begin{figure}
\centering
\rotatebox{0}{
\resizebox{7cm}{!}
{\includegraphics{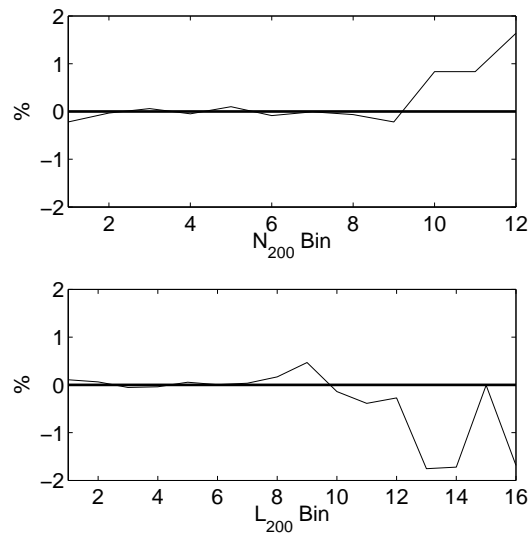}}
}
 \caption{The residual mass offset in the recovered masses of the spherical and triaxial halo samples for both luminosity (top panel) and richness (bottom panel) bins after correction by the effective spherical mass $M_{ES}$.  The triaxial bias is almost entirely removed in the low mass bins where the averaging over triaxial shape is most effective.  The quantity plotted is $(M_{3ax}-M_{Sph})/M_{Sph} - (\ave{M_{ES}} - \ave{M_{200}})/\ave{M_{200}}$.}
 \label{fig:plot3}
\end{figure}

\section{Lensing Analysis}\label{sec:lens}
We follow the binning procedure of \cite{johnston08} and bin the halos in 12 richness and 16 luminosity bins.  Within each bin we stack the lensed catalogues, assuming perfect knowledge of the cluster centre, and fit spherical NFW models to the high signal-to-noise stacked data extracted from within an annulus covering ${0}\arcminf{5}-{15}\arcminf{0}$ ($115-3500$ kpc) from the field centre to estimate the mean mass of each bin.

We primarily employ a radial shear profile fitting technique to fit the lensing masses.  Under this method the tangential ellipticities $\epsilon_t$ of the galaxies in the stacked catalogues are averaged in nine logarithmic radial bins, giving a mean radial shear profile.  This profile is then fit by minimising $\chi^2$
\begin{equation} \chi^2 = \sum_{i=1}^{N_{r_{bin}}}\frac{\left(\epsilon_{t,i} - g(r_i; \Pi)\right)^2}{\sigma_i^2}, \end{equation}
where $\epsilon_{t,i}$ is the mean tangential ellipticity in radial bin $i$, $g$ is the reduced shear calculated for the spherical NFW model being fit at the mean radius of the radial bin $r_i$ (for the lensing properties of spherical NFW halos see \cite{bartelmann96}), $\sigma_i$ is the statistical error on $\epsilon_{t,i}$, and $\Pi$ is a two-element vector of the parameters defining the model: virial mass $M_{200}$ and concentration $C$.  We assume a single effective mean lensing redshift of the background galaxies, which for our galaxy redshift distribution is $z_s=0.435$; for simplicity to isolate the effects of triaxiality we assume perfect knowledge of this value.   It is possible to carry out a more precise analysis by scaling the measured shear of each galaxy via its redshift to a single source plane prior to radial averaging, but because we are not specifically interested in the impacts of the redshift distribution in this paper, but aim to isolate the effects of triaxiality, we adopt the simpler approach of assuming a single mean effective lensing redshift.   The $\chi^2$ is minimised over $\Pi$ using a simple downhill minimiser from the GNU Scientific Library (GSL), and the values of $C$ and $M_{200}$ that minimise the function are returned as the mass and concentration estimate for the mass bin.  

To check that this method is robust (most especially our assumption that the lensed galaxies can be treated as if on a single plane at a single effective lensing redshift) we test our mass estimates using a maximum-likelihood method in which each galaxy in the stacked catalogue is treated individually.  We define the log-likelihood function in the standard manner for weak lensing following \cite{schneider00} and \cite{king01}
\begin{equation}\ell_{\gamma} = -\ln \mathcal{L} = -\sum_{i=1}^{n_{\gamma}}\ln  p_{\epsilon}(\epsilon_i|g(\vec{ \mathcal{\theta}_i}; \Pi))\end{equation}
where $\epsilon_i$ is the measured ellipticity of each galaxy, $n_{\gamma}$ is the number of galaxies used, the reduced shear $g$ is calculated for the spherical NFW model being fit, and $\Pi$ is again a two-element vector of the parameters $M_{200}$ and $C$.  We assign each lensed galaxy an individual redshift, and again to isolate the effects of triaxiality, assume perfect knowledge of those galaxy redshifts in our analysis.  The log-likelihood is minimised over $\Pi$ using the same GSL minimiser, and the maximum likelihood values for $C$ and $M_{200}$ are returned as the mass and concentration for the mass bin.  
\begin{figure*}
\centering
\rotatebox{0}{
\resizebox{14cm}{!}
{\includegraphics{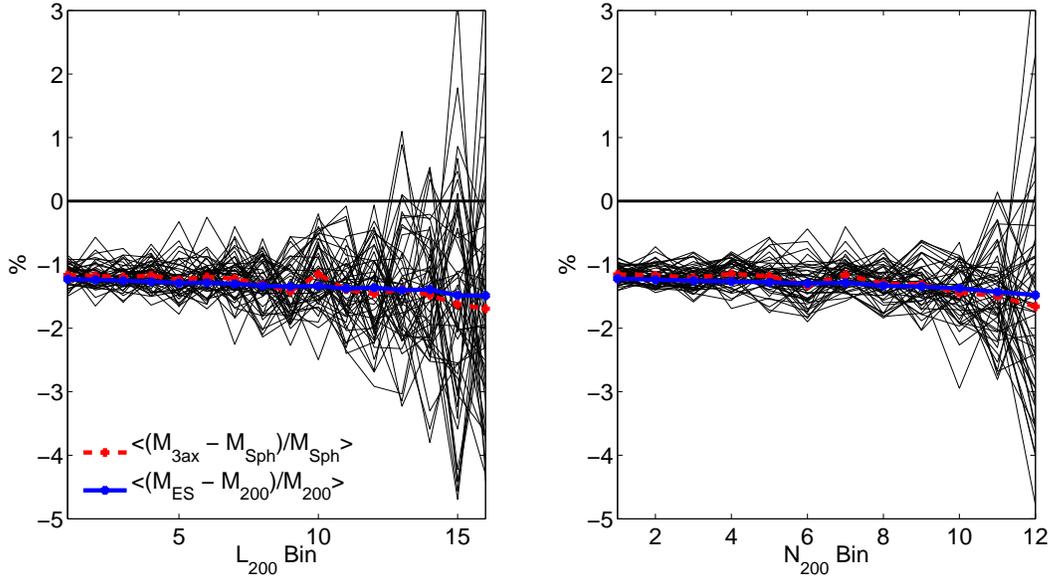}}
}
 \caption{Offsets in the masses recovered, averaged across 50 realizations of the lensing populations, for lensing by identical populations of 15,000 spherical and triaxial lensing halos, binned using SDSS observables.  The black lines plot the percent offset of the lensing mass recovered from the triaxial lensing population compared to that for the spherical population in each luminosity $L_{200}$ (left-hand panel) and richness $N_{200}$ (right-hand panel) bin, for all 50 realizations.  Overplotted in dashed red is the mean offset, and in solid blue is the mean value of the comparison of the true mean value of $M_{ES}$ to the true mean $M_{200}$ value for the triaxial population in each bin, each averaged across all 50 population realizations.}
 \label{fig:plot4}
\end{figure*}

Normalizing against this more complex method we confirm that the averaged radial shear method provides accurate mass estimates.  Because the radial shear method is much faster computationally than the maximum-likelihood method for large numbers of background galaxies (seconds vs. hours), we conduct the rest of our analysis using the averaged radial shear method only.

These methods are different to those employed in \cite{johnston08}, which employs a non-parametric cross-correlation technique, but are very similar to those used in other weak lensing mass reconstructions such as \cite{clowe06} and \cite{limousin07}.  We choose them for their simplicity and efficiency.  Because we fit our NFW lens halos with NFW models, assuming a parametric model will not induce any error or scatter in our calculations because of disagreement between the profile shapes of the model and data.

\section{Results and Discussion}\label{sec:results}
To begin, we simulate lensing by a single triaxial and corresponding spherical halo population (identical in all but shape) through 50 different realizations of the universe: i.e. different randomly chosen background galaxy populations.  To isolate the effects of triaxiality, in each realization the triaxial and spherical populations lens through identical universes: the background galaxy catalogues are independently and randomly chosen for each lens halo in each realization, but are identical between triaxial and spherical counterparts.  These 50 realizations of lensing through a single triaxial and corresponding spherical population will effectively average over most of the intrinsic galaxy shape noise, and so we expect that for the spherical case the recovered lensing masses will accurately recover the true bin values.  \cite{mandelbaum05} demonstrated that the masses recovered from a stacked lensing analysis in a universe with a steeply sloped mass function, such as that simulated here, are expected to fall somewhere below the mean and above the median mass of the true bin population.  If the mass dispersion within the bin is small, the difference between mean and median will be small, and that is indeed expected in this case in which we have included no intrinsic scatter in the mass-observable relations.

Figure \ref{fig:plot1} plots the offset in the recovered bin masses, averaged across the 50 realizations, compared to the {\it true} mean and median mass values ($\ave{M_{200}}$ and med$(M_{200})$) for the spherical and triaxial halo populations, in both the luminosity and richness bins. The error bars show the standard deviation of the 50 realizations.

To begin, we note that the masses of the spherical population (plotted with a solid black line) are recovered very well: indeed, the recovered masses tend to fall very slightly below the true mean values, and slightly above the true median values, exactly as expected from the Mandelbaum result.  The greater noisiness of the $L_{200}$ masses is as expected; because there are four more $L_{200}$ bins, the number of halos (and thus lensing signal) in each bin will decrease and so the uncertainty on each individual realization will be higher.  The higher levels of noise in the comparison to the median mass values is also expected, because the median is a noisier quantity than the mean.  Taken together, these indicate an acceptable recovery of the masses of the spherical population that follows the precedent of similar previous work; we are therefore confident to turn our attention to the recovery of the triaxial population.

\begin{figure*}
\centering
\rotatebox{0}{
\resizebox{14cm}{!}
{\includegraphics{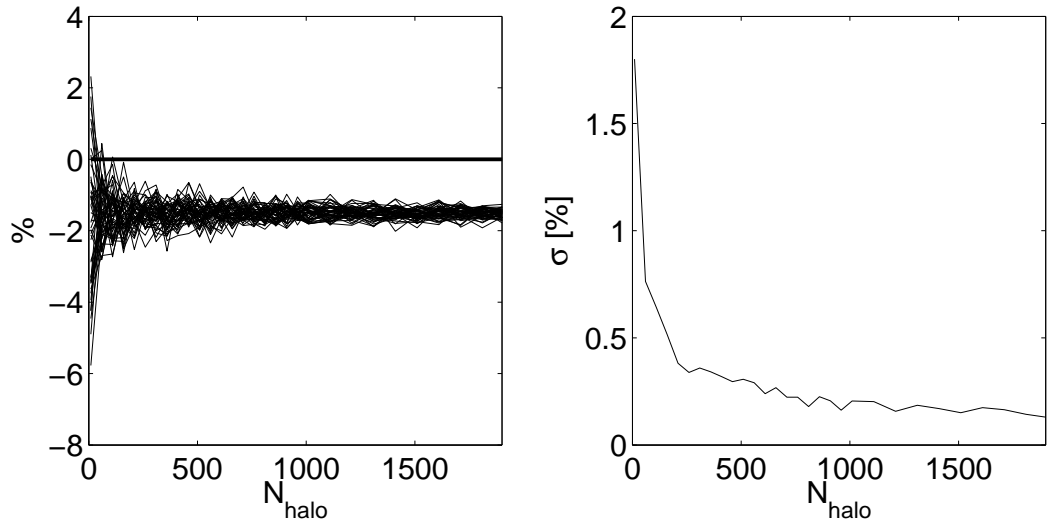}}
}
 \caption{Offsets in the masses recovered, averaged across 50 noise realizations, for lensing by identical populations of spherical and triaxial lensing halos in the 12th $N_{200}$ richness bin, plotted as a function of the number of halos in the bin, $N_{halo}$.  The left-hand panel plots the offset for each of the 50 realizations; the right-hand panel plots the dispersion around the mean deviation across the 50 realizations.  The lower the dispersion, the more effective the shape averaging and the better a spherical stacked lensing mass estimate will recover the {\it true} bin mass.}
 \label{fig:plot5}
\end{figure*}

\subsection{A triaxiality bias?}
The recovered masses for the triaxial population, averaged across the 50 realizations, are plotted with a blue dash-dot line in Figure \ref{fig:plot1}.  Crucially, the recovered mean mass values are consistently {\it 1-2 percent less} than those for their identical spherical counterparts.  By this offset, which is found consistently in every realization, the lensing mass recovered from triaxial halos is shown to be a biased estimator of $M_{200}$.  The noise between the spherical and triaxial populations is correlated because we have lensed through identical universes with lens populations identical in every way except for their shape -- thus the remnant galaxy shape noise is exactly the same for the spherical and triaxial cases.

Though the triaxial bias is well within the statistical error bars in all bins, it is still problematic.  Because the errors due to the shape noise of the lensing realizations are no longer centered on the true mass value -- but are instead consistently offset by a few percent -- comparisons between lensing constraints on the mass function and predictions from theory that assume symmetric or Gaussian errors will be incorrect.   In particular, if the mean masses measured by lensing are consistently low, as we have demonstrated here that they are, then the mass function will be directly affected because the masses with which $n(M)$ is calculated will be incorrect; this is crudely illustrated in Figure \ref{fig:plot1p5} in which the mean cumulative mass function $N(>M)$ is plotted for the 50 realizations of Figure \ref{fig:plot1} for both spherical and triaxial populations.  Note that this is independent of any change in the mass-observable relation used to sort halos into mass bins.  As better surveys with more background sources and smaller measurement errors come online, this bias will become even more important.  

Why this apparent difference when triaxial halos lens through an identical universe?  What physical or statistical explanation can be offered to explain why averaging does not work as expected in stacked lensing analyses?  Figure \ref{fig:plot2} shows that the way in which halo mass is defined is crucial in understanding this apparent triaxial bias.  The black lines plot, for each of the 50 realizations, the percent offset of the lensing mass recovered from the triaxial population $M_{3ax}$ compared to that recovered for the spherical population $M_{Sph}$
\begin{equation}\frac{M_{3ax} - M_{Sph}}{M_{Sph}},\end{equation}
in each luminosity $L_{200}$ (left-hand panel) and richness $N_{200}$ (right-hand panel) bin. As seen before in the average in Figure \ref{fig:plot1}, the lensing masses consistently underestimate the triaxial masses $M_{200}$ compared to those of their matched spherical counterparts by $\sim 1-2 \%$ in the low mass bins and even more in the high mass bins.  The very narrow dispersion between the 50 realizations illustrates that the scatter in the triaxial bias due to galaxy shape noise is far less than the bias itself, indicating the bias is robust in this respect.

Overplotted in the figure are comparisons of the true mean alternate masses -- calculated via the alternate mass measures listed in Section \ref{sec:mass} -- to the true mean triaxial mass for the triaxial population in each bin (the mass estimators are all equal for the spherical population): $(M_{alt} - M_{200})/M_{200}$.  It is hoped that the lensing mass recovered from the triaxial population is a good estimator of one of these alternate mass measures.  And indeed, the recovered lensing masses are seen to better estimate all three alternate mass measures.  Of the three, the 3D NFW mass $M_{3DNFW}$, plotted in green, provides the least improvement, as the lensing mass consistently overestimates it by a small percentage.

However, in the low mass bins the recovered lensing mass appear to be an excellent estimator of the effective spherical mass $M_{ES}$ (Eq. \ref{eq:MES}), and in the higher mass bins of the 2D NFW lensing mass $M_{2DNFW}$.  In those higher mass bins where $M_{2DNFW}$ is best recovered there are only a few halos ($< 100$) to average over; in these cases the 2D lensing mass estimator performs best because, though it still demands averaging over ellipticities in 2D to circular symmetry, it, by its 2-dimensional definition, does not require averaging over the unseen third dimension along the line of sight.  However, such a 2D mass estimator is of less general applicability because it is meaningful only in the context of lensing and requires the adoption of a line-of-sight orientation to be calculated.  Theory makes predictions for 3D mass functions -- and it is only through comparison to these predictions that a measured mass function can be leveraged to constrain cosmological parameters. Translating predictions to a derivative 2D mass is undesirable because their strength is diminished, though in problems in which the accurate estimate of the highest mass bins is critical those disadvantages may be outweighed by its accuracy.  When at all possible, accurate 3D masses are preferable.

Luckily, in the lower mass bins where the shape averaging is effective, the lensing mass performs even better as an estimator of the 3D effective spherical mass.  This is further illustrated in Figure \ref{fig:plot1}, where the lensing masses recovered from the triaxial population, averaged across the 50 realizations, are compared to the true effective spherical mean and median mass values ($\ave{M_{ES}}$ and med$(M_{ES})$).  In the low mass (many halo) bins, the errors in the stacked lensing masses recovered for the triaxial population exactly match those for the spherical population.  This is confirmed in Figure \ref{fig:plot3}, which shows the residual offset between the triaxial and spherical lensing masses when the true $M_{ES}$ values of the triaxial halo population are used to remove the triaxial bias; they are very close indeed to zero in the low mass bins where averaging is effective. We will return to the important question of when averaging over triaxial halo shape is effective in Section \ref{sec:av}

Now, to test the robustness of the apparent triaxial bias across different realizations of the lensing halo population, we perform lensing through 50 different halo populations, each chosen randomly from the distributions outlined in Sec. \ref{sec:sims}.  Again, identical triaxial and spherical populations are generated for each realization, and the background lensing catalogues are independently and randomly chosen for each lens, though matched between triaxial and spherical counterparts.  Figure \ref{fig:plot4} again plots the percent offset of the lensing mass recovered from the triaxial population compared to that for the counterpart spherical population for all 50 realizations, together with the comparisons of the alternate mass measures to $M_{200}$, as plotted in Figure \ref{fig:plot2}.  The triaxial bias is present as before, with greater scatter due to the noise introduced by the 50 different halo populations.  The scatter is very large in the high mass (few halo) bins where the averaging over triaxial shape is insufficient, because in these bins the shape distribution is quite different for every realization.  The scatter in these bins is symmetric, showing that the large underestimation of the mass seen in Figure \ref{fige:plot2} for the single lens population is only one possible outcome when the number of halos is too small; insufficient averaging can lead to both under- and over- estimates of the effective spherical mass.  Again, $M_{ES}$, whose average across the 50 realizations is plotted in blue in Figure \ref{fig:plot4}, is the mass measure best estimated by the recovered lensing mass, as long as the averaging is sufficient.  Our findings are therefore shown to be robust both for different realizations of the lensing universe and of the lensing population.

Though previous work in \cite{corless08} has shown averaging over spherical models individually fit to weak lensing data to give approximately correct mean values for triaxial mass $M_{200}$ and concentration over a population of 100 equal mass triaxial halos, it was not concerned with errors of a few percent.  Those finding are therefore consistent with those in this paper, in which the averaging problem is examined in far greater detail.

Though the bias seen in the recovery of the triaxial mass function is small, at this time significantly smaller than other sources of error (crucially in the normalisation and scatter in the mass-observable relationships), it is important.  In future surveys such as those reviewed by the DETF its relative weight compared to other errors will increase as those other sources of error are controlled. Even now, a bias of any size that consistently underestimates mass can be important given the exponential dependence of the normalization of the matter power spectrum on the high-mass end of the mass function \citep{albrecht06}; given that the bias is easily removed, it should be.

We have demonstrated that the presence of triaxial halos in a lensing population can affect the measurement of average masses in bins typical of those used in stacked analyses to measure the mass function.  The choice of mass becomes important, as the presence of triaxiality breaks the degeneracy between many possible mass definitions.  We showed that the 2D lensing mass is best recovered when shape averaging is not adequate, but that the 3D effective spherical mass is best when there are enough halos in a given bin to make the averaging over triaxial shape effective.  

\subsection{Effective Triaxial Averaging}\label{sec:av}
But how many halos are required for that to be so?  To test this, we take a single mass bin -- the 12th richness bin, at the uppermost end of the mass function -- and populate it with an increasing number of halos $N$.  We do this for 50 realizations of paired triaxial and spherical halo populations.  We are looking for the number of halos at which the 'triaxial bias' becomes constant -- i.e. the relationship between the recovered masses for the triaxial and spherical populations is constant, such that it can easily be accounted for by using $M_{ES}$.  The left-hand panel of Figure \ref{fig:plot5} plots the percent deviation between the triaxial and spherical recovered masses as a function of $N$ for all realizations, and the right-hand panel plots the dispersion of those deviations around the mean over the 50 realizations.  As expected, the dispersion is largest when the number of halos in the bin is smallest.  It decreases significantly to about $0.3\%$ by N=500, and flattens out to $0.2\%$ by $N=1600$.  This plateau value represents the dispersion due to a combination of the noise of the different realizations of the lensing population and the intrinsic galaxy shape noise.

When designing strategies for present and future surveys, it is important to ensure there are an adequate number of halos in each mass bin for averaging over triaxial shape.  Our results suggest 100 halos as a minimum number, and 500 as highly preferable. For the upper reaches of the mass function where such numbers will not be available, individual triaxial mass modelling using techniques such as those in \cite{corless08} may be necessary, as the approximation of average sphericity no longer holds.  Neglecting this caution may result in errors of up to $\sim 5\%$ in virial mass estimates.

\subsection{Normalization and Scatter in the Mass-Observable Relations}
\begin{figure}
\centering
\rotatebox{0}{
\resizebox{8cm}{!}
{\includegraphics{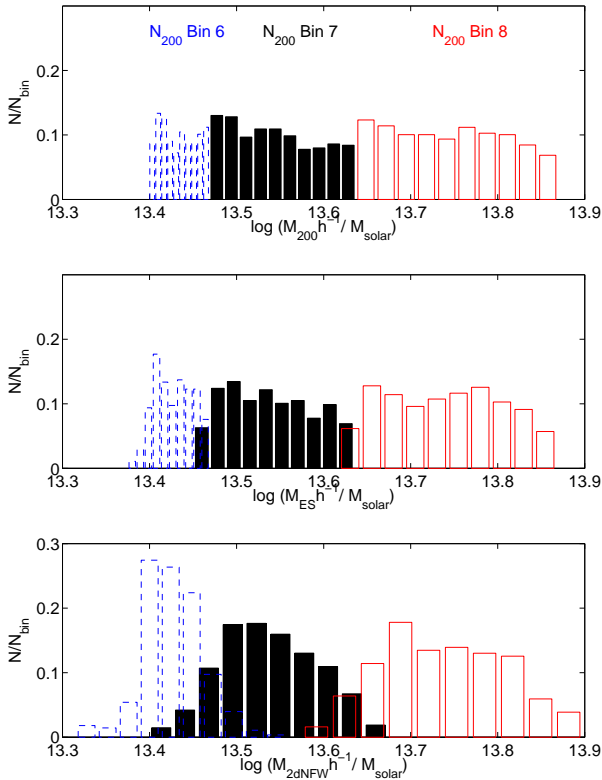}}
}
 \caption{Distributions of triaxial masses $M_{200}$ (top panel), 3D effective spherical masses $M_{ES}$ (middle panel), and 2D circular lensing masses $M_{2DNFW}$ (bottom panel) of the halos in the 6th-8th richness bins for a single triaxial population.  The standard deviation within each bin increases by $1-5\%$ if the lensing mass $M_{2DNFW}$ is used rather than the triaxial or effective spherical masses.}
 \label{fig:plot6}
\end{figure}

In addition to introducing a potential bias in direct estimates of masses from statistical weak lensing, the presence of triaxial structure in dark matter halos also can introduce an effective scatter into the mass-observable relations that are used to sort halos into bins for stacked analysis.  To illustrate this, Figure \ref{fig:plot6} plots the bin members of the 6th, 7th, and 8th richness bins for a single triaxial halo population.  The top panel plots them as function of $M_{200}$, the middle as a function of $M_{ES}$, and the bottom panel as a function of $M_{2DNFW}$.  For a spherical population all of these would be identical to the $M_{200}$ distribution (note that we neglect all intrinsic mass-observable scatter).  

In this simulation we have assumed that the true mass-observable relation is between the observable and the triaxial halo mass $M_{200}$ when sorting halos into mass bins.  This may not be the case; perhaps the tightest mass-observable relation with least scatter utilizes the effective spherical mass, or some other mass measure.  However, what this figure clearly illustrates is the critical role consistent and well-understood mass estimators play in accurately characterizing the mass-observable relations.  Quantitatively, the scatter $\sigma_M$ in the mass-richness and mass-luminosity relations using the the mean effective lensing mass  ($M_{2DNFW}$) varies from about $6\%$ in low mass bins to over $8\%$ in high mass bins, compared to no scatter in the triaxial mass $M_{200}$.  For $M_{ES}$, the effective scatter is $\sim 2\%$.

As an example, determining the scatter in the mass-observable relations by modelling individual galaxy clusters using a standard spherical lensing model, which gives an estimate of $M_{2DNFW}$, will almost certainly significantly increase the effective scatter in the  mass-observable relation.  this is because there is little reason to expect richness or luminosity (or X-ray luminosity or temperature) to be strongly dependent on lensing orientation, and so comparing these quantities with 2D lensing masses will necessarily increase the scatter in the apparent mass-observable relations.  Such scatter is widely known to be important (\cite{albrecht06}; \cite{voit04}) because the mass function is steeply sloped, such that there are more low mass halos to scatter up than high mass halos to scatter down; this causes the mass recovered for a bin to decrease as the scatter increases.  The scatter among bins must be well known in order to correct for this effect and accurately interpret the mass measurement in each bin.  

Regarding the mass-observable normalizations, if a stacked lensing analysis such as this is used to determine the normalization of the mass-observable relation, it is crucial to recognize that it is the relationships of observables to $M_{ES}$ that are being characterized, and not those to $M_{200}$ or any other mass estimator.  If care is not taken to identify and account for the 3D mass estimator that the lensing analysis accurately recovers, future studies that use the mass-observable relations derived from lensing to constrain the mass function directly from observables without any independent measure of mass will suffer from the same mass biases of the weak lensing analysis.


At present, the increase in mass-observable scatter due to triaxiality, even using the poor $M_{2DNFW}$ indicator, is small compared to the best estimates of the intrinsic scatter in most mass-observable relations.  Recent work puts the intrinsic scatter in the X-ray mass-temperature and mass-luminosity relations at $\sim 15\%$ and  $45\%$ respectively (\cite{vikhlinin06}; \cite{stanek06}), and that in the mass-richness relation perhaps as high (the relation between richness and X-ray luminosity is found by \cite{kochanek03} to have scatter of 0.33 dex, greater than $100\%$ for some values of $L_x$).  Most optimistically, the relation between $M_{500}$, a mass measure frequently used in X-ray analyses, and $Y_X$, a proxy obtained by multiplying together the gas mass and X-ray temperature was shown in \cite{kravtsov06} to have an intrinsic scatter of less than $10\%$.  Thus, in present analyses using all but the last relation -- which is unavailable as an observable for large optical weak lensing surveys unless they are accompanied by extensive complementary X-ray observations -- the scatter due to triaxiality is non-negligible but unlikely to be of primary concern.  However, in future surveys in which the intrinsic scatter in the mass-observable relations can be controlled to a few percent, as the DETF argues it must be, then the scatter due to triaxiality must be carefully considered and controlled.

\subsection{Summary}
We investigate the impacts of triaxiality on the accuracy of mass estimates derived from stacked weak lensing data with the aim of constraining the cluster mass function.   Though the quantitative results of our study may not directly translate to other analysis methods using non-parametric reconstructions or cross-correlation stacking techniques, the general finding will hold true: the presence of triaxiality makes the choice of mass estimator crucially important.

Most importantly, we find that triaxiality can introduce a bias in the masses recovered from a triaxial halo population using a stacked lensing analysis, giving mass values that are a few percent low in all bins compared to those of their spherical counterparts.    This bias in the mean masses statistically measured by lensing will lead directly to incorrect constraints on the mass function $n(M)$.  Further, calibrations of the scatter in, and the normalization of, the mass-observable relations critical to accurate measurement of the mass function derived from lensing will also be made inaccurate by the bias.  However, we show that this bias is removed if the effective spherical mass $M_{ES}$, rather than the triaxial virial mass $M_{200}$, is employed in defining the triaxial halo masses.  While the effective spherical mass $M_{ES}$ is not the ideal mass measure for all problems -- especially those studying individual clusters (see \cite{corless08})-- and has the disadvantage of treating dark matter halos as approximately spherical when all predictions suggest a signficantly triaxial population, it is most suitable in this case. Further, we demonstrate that least 100, and preferably 500, halos are needed in every mass bin to make averaging over triaxial shape effective and so reliably recover the effective spherical mass $M_{ES}$.  

Thus, stacked lensing analyses aiming to constrain the mass function or characterize the mass-observable relations provide accurate mass estimates {\it only if the 3D mass estimators employed are very carefully selected} and an adequate number of halos are stacked in each observable bin. With these caveats, applied by carefully calibrating any stacked lensing mass reconstruction method with simulations, they can be applied to future weak lensing surveys such as JDEM, LSST, and DES unhindered by galaxy cluster and group triaxiality.

\section*{Acknowledgements}
This work was supported by the National Science Foundation, the Marshall Foundation, and the Cambridge Overseas Trust (VLC) and the Royal Society (LJK).  We thank Antony Lewis for helpful discussions, and an anonymous referee for a careful review that helped to significantly clarify our arguments.







\bsp

\label{lastpage}


\begin{thebibliography}{99}
\bibitem[\protect\citeauthoryear{Albrecht et al.}{2006}]{albrecht06} Albrecht, a. et al. 2006, {\it Report of the Dark Energy Task Force}, astro-ph/0609591
\bibitem[\protect\citeauthoryear{Bartelmann}{1996}]{bartelmann96} Bartelmann, M. 1996, A\&A, 313, 697, astro-ph/9603101
\bibitem[\protect\citeauthoryear{Bahcall et al.}{2003}]{bahcall03} Bahcall, N.A., et al. 2003, ApJ Supplement, 148, 243
\bibitem[\protect\citeauthoryear{Bett et al.}{2007}]{bett07} Bett, P., Eke, V., Frenk, C.S., Jenkins, A., Helly, J., Navarro, J. 2007, astro-ph/0608607
\bibitem[\protect\citeauthoryear{Canizares}{1982}]{canizares82} Canizares, C. R. 1982, ApJ, 263, 508
\bibitem[\protect\citeauthoryear{Clowe et al.}{2006}]{clowe06} Clowe, D., et al. 2006, A\&A, 451, 395
\bibitem[\protect\citeauthoryear{Corless \& King}{2007}]{corless07} Corless, V. \& King, L., 2007, MNRAS, 380, 149, astro-ph/061191
\bibitem[\protect\citeauthoryear{Corless \& King}{2008}]{corless08} Corless, V. \& King, L., 2008, Accepted for publication in MNRAS
\bibitem[\protect\citeauthoryear{Dodelson}{2004}]{dodelson04} Dodelson, S. 2004, PhRvD, 70, 023008, astro-ph/0309277
\bibitem[\protect\citeauthoryear{Evrard et al.}{2002}]{evrard02} Evrard, A.E.et al., 2002, ApJ, 573, 7
\bibitem[\protect\citeauthoryear{Hoekstra}{2003}]{hoekstra03} Hoekstra, H. 2003. MNRAS, 339, 1155, astro-ph/0208351
\bibitem[\protect\citeauthoryear{Jing \& Suto}{2002}]{jing02} Jing, Y. P., \& Suto, Y. 2002, ApJ, 574, 538, astro-ph/0202064
\bibitem[\protect\citeauthoryear{Johnston et al.}{2008}]{johnston08} Johnston, D.E., et al., 2008, astro-ph/0709.1159
\bibitem[\protect\citeauthoryear{King \& Schneider}{2001}]{king01} King, L. J., \& Schneider, P. 2001, A\&A, 369, 1, astro-ph/0012202
\bibitem[\protect\citeauthoryear{Keeton}{2001}]{keeton01} Keeton, C.R. 2001, astro-ph/0102341
\bibitem[\protect\citeauthoryear{Kochanek et al.}{2003}]{kochanek03} Kochanek, C., White, M., Huchra, J., Macri, L., Jarrett, T.H., Schneider, S.E., Mader, J. 2003, ApJ, 585, 161
\bibitem[\protect\citeauthoryear{Koester et al.}{2007}]{koester07} Koester, B. P. et al. 2007, ApJ, 660, 239
\bibitem[\protect\citeauthoryear{Kravtsov, Vikhlinin, \& Nagai}{2006}]{kravtsov06} Kravtsov, A.V., Vikhlinin, A., \& Nagai, D. 2006, ApJ, 650, 128
\bibitem[\protect\citeauthoryear{Limousin et al.}{2007}]{limousin07} Limousin, M., et al. 2007, ApJ, 668, 643 
\bibitem[\protect\citeauthoryear{Mandelbaum et al.}{2005}]{mandelbaum05}Mandelbaum, R., Tasitsiomi, A., Seljak, U., Kravtsov, A. V., Wechsler, R. H. 2005, MNRAS, 362, 1451
\bibitem[\protect\citeauthoryear{Mandelbaum \& Seljak}{2007}]{mandelbaum07}Mandelbaum, R. \& Seljak, U. 2007, JCAP, 6, 24
\bibitem[\protect\citeauthoryear{Meneghetti et al.}{2007}]{meneghetti07}Meneghetti, M., Bartelmann, M., Jenkins, A., Frenk, C. 2007, MNRAS, 381, 171
\bibitem[\protect\citeauthoryear{Navarro et al.}{1997}]{navarro97}Navarro, J. F., Frenk, C. S., White, S. D. M., 1997, ApJ, 490, 493, astro-ph/9611107
\bibitem[\protect\citeauthoryear{Neto et al.}{2007}]{neto07} Neto, A.F., et al., 2007, MNRAS, 381, 1450
\bibitem[\protect\citeauthoryear{Oguri, Lee, \& Suto}{2003}]{oguri03} Oguri, M., Lee, J., Suto, Y. 2003, ApJ, 599, 7, astro-ph/0306102
\bibitem[\protect\citeauthoryear{Oguri et al.}{2005}]{oguri05} Oguri, M., Takada, M., Umetsu, K., Broadhurst, T. 2005, ApJ, 632, 841
\bibitem[\protect\citeauthoryear{Oyaizu et al.}{2007}]{oyaizu07} Oyaizu, H., Lima, M., Cunha, C., Lin, H., Frieman, J. 2007, submitted to ApJ, astro-ph/0711.0962
\bibitem[\protect\citeauthoryear{Reiprich \& Boehringer}{2002}]{reiprich02} Reiprich, T.H. \& Boehringer, H. 2002, ApJ, 567, 716
\bibitem[\protect\citeauthoryear{Schneider, King, \& Erben}{2000}]{schneider00} Schneider, P., King, L. J., Erben, T. 2000, A\&A, 353, 41, astro-ph/9907143
\bibitem[\protect\citeauthoryear{Shaw et al.}{2006}]{shaw06} Shaw, L.D., Weller, J., Ostriker, J.P., Bode, P. 2006, ApJ, 646, 815, astro-ph/0509856
\bibitem[\protect\citeauthoryear{Sheth, Mo, \& Tormen}{2001}]{sheth01} Sheth, R.K., Mo, H.J, Tormen, G. 2001, MNRAS, 323, 1, astro-ph/9907024
\bibitem[\protect\citeauthoryear{Stanek et al.}{2006}]{stanek06} Stanek, R., Evrard, A.E., Boehringer, H., Schueker, P., Nord, B. 2006, ApJ, 648, 956
\bibitem[\protect\citeauthoryear{Vikhlinin et al.}{2006}]{vikhlinin06} Vikhlinin, A., Kravtsov, A., Forman, W., Jones, C., Markevitch, M., Murray, S., Van Speybroeck, L. 2006, ApJ, 640, 691
\bibitem[\protect\citeauthoryear{Vikhlinin et al.}{2008}]{vikhlinin08} Vikhlinin, A. et al. 2008, submitted to ApJ, astro-ph/0805.2207
\bibitem[\protect\citeauthoryear{Voit}{2005}]{voit04} Voit, G. M. 2005, Reviews of Modern Physics, 77, 207
\bibitem[\protect\citeauthoryear{York et al.}{2000}]{york00} York, D.G., et al., 2000, AJ, 120, 1579
\end{thebibliography}
\end{document}